# Frequency modulated few-cycle optical pulse trains induced controllable ultrafast coherent population oscillations in two-level atomic systems

Parvendra Kumar and Amarendra K. Sarma*

Department of Physics, Indian Institute of Technology Guwahati, Guwahati-781039, Assam, India
*E-mail: aksarma@iitg.ernet.in

We report a study on the ultrafast coherent population oscillations (UCPO) in two level atoms induced by the frequency modulated few-cycle optical pulse train. The phenomenon of UCPO is investigated by numerically solving the optical Bloch equations beyond the rotating wave approximation. We demonstrate that the quantum state of the atoms and the frequency of UCPO may be controlled by controlling the number of pulses in the pulse trains and the pulse repetition time respectively. Moreover, the robustness of the population inversion against the variation of the laser pulse parameters is also investigated. The proposed scheme may be useful for the creation of atoms in selected quantum state for desired time duration and may have potential applications in ultrafast optical switching.

Recent progress in the generation of well controlled shaped-few-cycle optical pulses with durations of only a few periods of the optical radiation gives a new boost to the study of the so-called light-matter interaction [1-4]. With the advent of these few-cycle pulses, the research in quantum coherent control is getting tremendous attention owing to high peak powers, enormous spectral bandwidth and selectivity offered by the few-cycle pulses [5-7]. In particular, controlling the population transfer among the quantum states of atoms has remained one of the foremost areas of research in optical physics. This is primarily due to many well-known potential applications including collision-dynamics, atomic interferometry, spectroscopy and optical control of chemical reactions etc. In the context of coherent population transfer, several schemes, such as stimulated Raman adiabatic passage (STIRAP) [8-10], adiabatic rapid passage (ARP) [11], Raman chirped adiabatic passage (RCAP) [12] and temporal coherent control (TCC) [13] etc., have been proposed and exploited by many authors. Recently, it has been demonstrated both experimentally and theoretically that femtosecond pulses and train of femtosecond pulses may be used for efficient and robust controlling of the population transfer among the quantum states of atoms [14-16]. Recently, coherent population oscillations (CPO) is studied by many authors in $\Lambda$-like three level atoms, in the context of electromagnetically induced transparency (EIT) [17], spatial optical memory [18], superluminal light [19] and ultraslow light [20]. The present brief report is largely motivated by the work of M. Scalora et al. [21] and M. E. Crenshaw et al. [22], where authors investigated that the coherent population oscillations (CPO) in two level atomic systems as induced by varying the Rabi frequency of the interacting pulse. In this work, we report a scheme of near complete population transfer with the trains of few cycle pulses in which the atoms may be put on hold in selected state for desired time duration with the judicious choice of pulse parameters. The phenomenon of coherent population oscillations (CPO) induced by the train of few-cycle pulse is investigated. This new kind of CPO is termed as ultrafast coherent population oscillations (UCPO) owing to the ultrafast nature of the oscillations. The novel feature of the reported UCPO is that the frequency of oscillations may be controlled by controlling the pulse repetition time. The model atomic system for this work is chosen to be a two level Na atom. However, the proposed scheme may be applicable to all other atoms, which could be modelled as two level atoms. It may be noted that, recently, it has been pointed out by many authors that the so-called rotating wave approximation (RWA) do not hold when one deals with few-cycle pulse related phenomena and should work in the non-RWA regime [16, 23, 24]. Hence, in this work we are working in the non-RWA regime and assume that all the atomic relaxation times are considerably longer than the interaction time with the laser pulses.

Our analysis is based on the scheme depicted in Fig.1. In Fig.1 (a), the states $|1\rangle$ and $|2\rangle$ refers to $3s$ and $3p$ quantum states of neutral sodium atoms respectively.

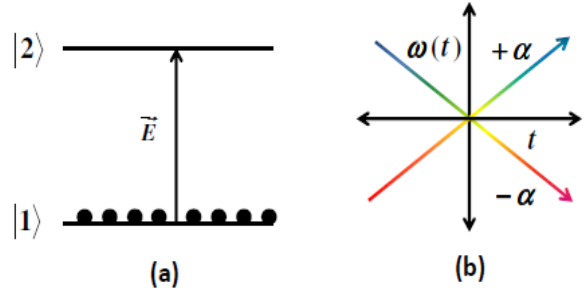

Fig.1 (Color online) (a) Two level atomic system (b) Time dependent frequency of up-chirped (+$\alpha$) and down-chirped (-$\alpha$) pulses.

The total electric field for the train of pulse can be written as $\vec{E}(t-nt_r) = \sum_{n=0}^{N-1} \hat{\varepsilon} E_0 f(t-nt_r)\cos(\omega(t-nt_r)+\alpha(t-nt_r)^2)$. Here, $f(t-nt_r) = \exp\left(-\left((t-nt_r)/\tau\right)^2\right)$ is the Gaussian shaped pulse envelope, $\tau = 1.177\tau_p$, where $\tau_p$ is the temporal pulse width at full width at half maximum (FWHM), N is the number of pulses, $t_r, \alpha$ and $\hat{\varepsilon}$ are the pulse repetition time, the chirp rate and the electric field polarization direction respectively. In this work, the states $|1\rangle$ and $|2\rangle$ refer to the down and the up state respectively. The optical Bloch equations, without invoking the so called rotating wave approximation, describing the temporal evolution of the density matrix elements, are:

$$\frac{du}{dt} = \Omega v - \frac{u}{T_2}$$
$$\frac{dv}{dt} = -\Omega u - 2\Omega_R(t)w - \frac{v}{T_2} \quad (1)$$
$$\frac{dw}{dt} = 2\Omega_R(t)v - \frac{(w+1)}{T_1}$$

Here $u, v$ and $w$ (population inversion) are the three components of the Bloch vector. $T_2$ and $T_1$ are respectively the dipole-dephasing and spontaneous decay time, $\Omega$ is the transition frequency of the two level atoms and $\Omega_R(t)$ is the Rabi frequency, defined as $\Omega_R(t) = \vec{\mu}.\vec{E}(t)/\hbar$. In the present study, we have neglected the terms associated with $T_2$ and $T_1$. This may be attributed to the fact that the atom-field interaction time, owing to extremely short duration of the few-cycle laser field, is negligibly small compared to $T_2$ and $T_1$.

We solve Eq. (1) numerically using a standard fourth-order Runge-Kutta method. We assume that initially all the atoms are in the ground state $|1\rangle$. We use the following typical parameters: $\Omega = \omega = 3.19$ rad/fs, $\tau = 25$ fs and $\mu_{12} = 1.85 \times 10^{-29}$ Cm [25], peak Rabi frequency, $\Omega_{21} = \mu E_0/\hbar = 0.50$ rad/fs and $\alpha = 0.015$ fs$^{-2}$. Fig. 2 depicts the temporal evolution of the population inversion with respect to the pulse repetition time.

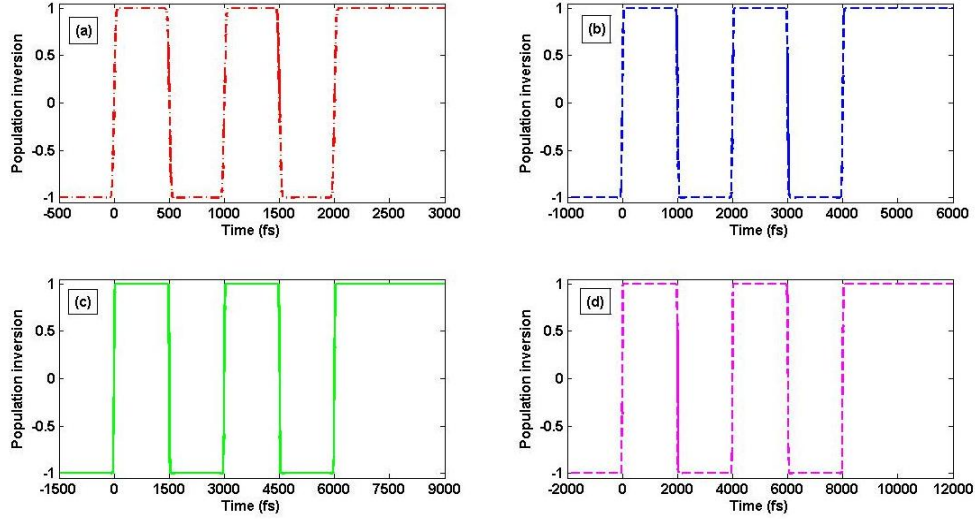

Fig. 2 (Color online) Temporal evolution of population inversion for N=5, with pulse repetition time, (a) $t_r$ =500 fs (b) $t_r$ =1000 fs (c) $t_r$ =1500 fs and (d) $t_r$ =2000 fs

It could be observed from Fig. 2 (a) that the all the atoms are transferred to up-state during the interaction with initial pulse (n=0) in the pulse train with peak Rabi frequency at t=0. On the other hand, all the atoms are transferred to down-state, when the next pulse (n=1) in the pulse train interact with the up-state atoms. The holding time of atoms in the up-state is nearly equal to the pulse repetition time ($t_r$), chosen to be 500 fs in Fig. 2(a). Moreover, it could be seen from Fig. 2 that every odd number of pulses act as a pump pulse, leads to the atoms in the up-state while every even number of pulses act as a dump pulse, leads to the atoms in the down-state. Therefore, the final state of atoms may be controlled by manipulating the number of pulses in the pulse train. A careful inspection of Fig. 2 reveals that the hold on time of atoms ($\tau_a$) in up and down-states follows the relation, $\tau_a \approx t_r$ and the frequency of ultrafast coherent population oscillations ($f_{ucpo}$) between up and down state is $f_{ucpo} \approx 1/2t_r$. Hence the hold on time of atoms in down, up-states and the frequency of population oscillations may be controlled by just controlling the pulse repetition time. Physically speaking, the pumping and dumping of atoms to the up and the down-states occurs via the so-called stimulated absorption and stimulated emission respectively as the spontaneous processes may not take place at such a short interaction time scale. It may be noted from Fig.2 that the system exhibits a step like transition, from absorbing (w<0) to amplifying (w>0) and amplifying (w>0) to absorbing (w<0) as a function of the number of the pulses in the pulse train. The atomic medium, for odd number of pulses in the pulse train may be interpreted as an 'on' state and for the even number of pulses in the pulse train may be interpreted as an 'off' state.Therefore the present scheme may serve as a unique ultrafast optical switch, in which switching time may be controlled as follows: when a switching signal enters into the absorbing medium (w<0), the interaction may take place between the switching signal and the absorbing medium; thereby the switching signal may get absorbed in the medium due to the absorptive character of the medium. This corresponds to the 'off' state of the optical switch for a switching signal. On the other hand, when a switching signal enters into the amplifying medium (w>0), the interaction may take place between the switching signal and amplifying medium; thereby the switching signal may not get absorbed in the medium due to the amplifying character of the medium. This corresponds to the 'on' state of optical switch for a switching signal. It is worthwhile to mention that the reported UCPO is almost similar to the one reported by Scalora et al. [21, 22] for ultrafast optical switching.

Therefore the present scheme may also find similar applications. However, in the present work we are considering a dilute atomic medium and hence the effect of near-dipole-dipole (NDD) interaction is not taken into account. In order to verify the robustness of the scheme, in Fig. 3, we depict the evolution of $w(\infty)$ against the variation of chirp rate for $t_r = 1000$ fs and N=5.

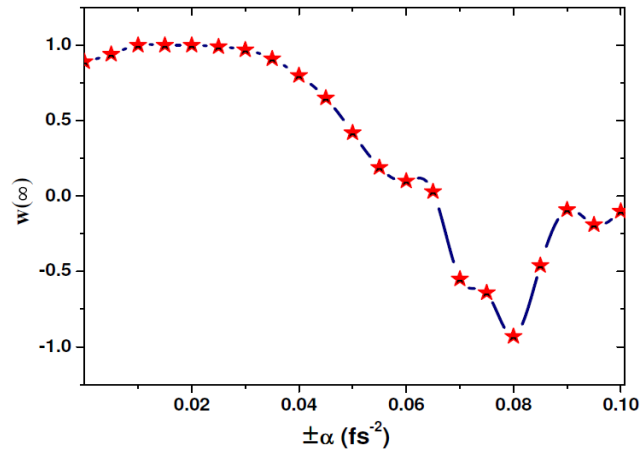

Fig.3 (Color online) Evolution of $w(\infty)$ against the variation of chirp rate ($\alpha$).

It is clear from Fig. 3 that the final population inversion is sufficiently robust against the variation of the chirp rate for chosen $t_r$ and N. From numerical study, the final population inversion is found to be robust against the variation of pulse duration and Rabi frequency as well. The final population transfer is found to be robust against the variation in chirp rate, pulse duration and Rabi frequency to a sufficiently large range e.g. $\alpha \approx \pm 0.01 - \pm 0.03\, \text{fs}^{-2}$, $\tau_p \approx 19 - 110\, \text{fs}$ and $\Omega_{12} \approx 0.35 - 1.50$ rad/fs respectively. The remarkable features, namely, $\tau_a \approx t_r$ and $f_{ucpo} \approx 1/2t_r$, of the present study remains invariant to the pulse envelopes of various shapes such as: Sech, Sinc and Lorentz shaped pulse envelope. Since the final population inversion is sufficiently robust against the variation of the laser pulse parameters, the proposed scheme may enable efficient generation of complete population inversion in atoms of an ensemble located in different spatial points covered by the laser pulse. Hence the proposed scheme may be explored experimentally as well. The scheme may also be exploited to measure pulse repetition rate.

To conclude, we have demonstrated the phenomenon of controllable UCPO in the two level atoms by utilizing the train of frequency modulated few-cycle pulses. It is shown that the hold on time for atoms in the up and the down-states may be controlled by controlling the pulse repetition time. The possible applications of the proposed scheme in ultrafast optical switching are also explored. The final population transfer to the target quantum state is found to be sufficiently robust against the variation of the Rabi frequencies and the chirp rates. The proposed scheme may find new applications in the area of ultrafast optical switching.